# A Novel Structure for Double Negative NIMs towards UV Spectrum with High FOM

**Jianwei Tang[1] and Sailing He[1,2*]**

[1] *Centre for Optical and Electromagnetic Research, JORCEP[KTH-ZJU Joint Research Center of Photonics], Zhejiang University (ZJU), Hangzhou 310058, China*
[2] *Department of Electromagnetic Engineering, School of Electrical Engineering, Royal Institute of Technology (KTH), 100 44 Stockholm, Sweden*
[*]*sailing@kth.se*

**Abstract:** A novel ring structure is proposed for double negative NIMs at visible light spectrum with high FOM (e.g. about 11 at a wavelength of 583 nm) and low loss. Besides the effective medium theory, an equivalent circuit model is also given to explain physically why our novel structure can give double negative behavior with low loss. Adapted from the original ring structure, two other types of structures, namely, disk and nanowire structures, are also given to further push double negative NIMs toward ultraviolet (UV) spectrum.

## 1. Introduction

About forty years ago Veselago theoretically investigated a negative refractive index medium by assuming both permittivity and permeability are negative [1]. However, such a negative-index material (NIM) stayed in theoretical assumption until its first demonstration in microwave frequency range using the so-called split ring resonators (SRRs) combined with thin metallic wires [2]. Later, SRRs have been successfully scaled to terahertz and even infrared frequency ranges [3, 4]. Nevertheless, the magnetic response of SRRs will saturate when scaling to an optical wavelength due to the kinetic inductance of a metal [5]. The so-called "fishnet" NIM structure can be easily scaled to some optical wavelengths [6], and most of the optical NIMs were realized with "fishnet" structures [7-13].

The loss of an NIM is indicated by figure of merit ($FOM = -\text{Re}(n)/\text{Im}(n)$). To obtain an NIM, condition $\varepsilon'\mu''+\mu'\varepsilon''<0$ suffices (no need for both permittivity and permeability to be negative). However, if high FOM (i.e., low loss) is desired, double negative NIMs are in general better than single negative NIMs.

The shortest wavelength at which double negative NIM (through the effective medium theory) has been reported to date is 813nm with a FOM of 1.3 [11] and 725nm (red light) with a FOM of 1.05 [12], using a fishnet structure. It seems quite difficult to reduce further the wavelength due to the quickly increasing loss and decreasing absolute value of the negative real part of the permittivity of silver (whose loss is the lowest among metals). The working frequency of single negative NIMs has just been pushed to yellow light, however, the FOM is quite low [13]. So far there is no report on UV NIM. In this letter, we proposed a novel structure for double negative NIMs of good quality at visible light spectrum and even UV spectrum, with much higher FOM and lower loss as compared with any fishnet structure. This opens up the green, blue, violet and even UV light spectrum for a double negative NIM of low loss, which has important implications of e.g. high resolution lithograph (through a superlens). The proposed structure consists of arrays of stacks of silver rings in e.g. SiO$_2$ (with a relative permittivity of 2.25) matrix as shown in Fig. 1(a). Numerical simulations were performed using CST Microwave Studio. The permittivity of silver was obtained from experimental data [14]. The effective electromagnetic parameters of the structure were retrieved from some unit cell simulation results [15, 16]. The effective refractive index, permittivity and permeability are shown in Fig. 1(b) and (c). In the wavelength region from 538 nm to 598 nm, both the effective permittivity and permeability are negative, indicating double negative NIM, with the largest FOM of about 11 at 583 nm. To confirm that the correct branch of the real part of the refractive index has been chosen during the standard retrieval procedure, different number of layers of unit cells were simulated and good convergence was observed, as shown in Fig. 1(b). For comparison with some experimental realizations of fishnet NIMs [11-13], the FOM is also shown in Fig. 1(b) in the grey solid line when the silver loss increases by a factor of three [13] as compared to the loss of bulk silver [14] in order to account for the surface roughness caused by experimental imperfection. Clearly, our structure exhibits much higher FOM (even one order higher) than fishnet structures operating in visible light [13]. We choose different branches for different wavelength regions to meet the requirement that the electromagnetic parameters should be consistent for different number of layers of unit cells. This causes the discontinuities of the curves in Fig. 1(b) and (c). Such a discontinuity is common in metamaterials [17, 18] and the origin has been well studied in [17].

In order to double check the negative refraction behavior, we arrange the NIM structure into a prism shape to demonstrate numerically the Snell refraction [2]. The refractive index "measured" through this numerical simulation according to Snell's law was compared with that retrieved from S parameters. They are found to be in good agreement, as shown in Fig. 1(b). The negative refraction behavior at 555 nm wavelength is shown in Fig. 1(d).

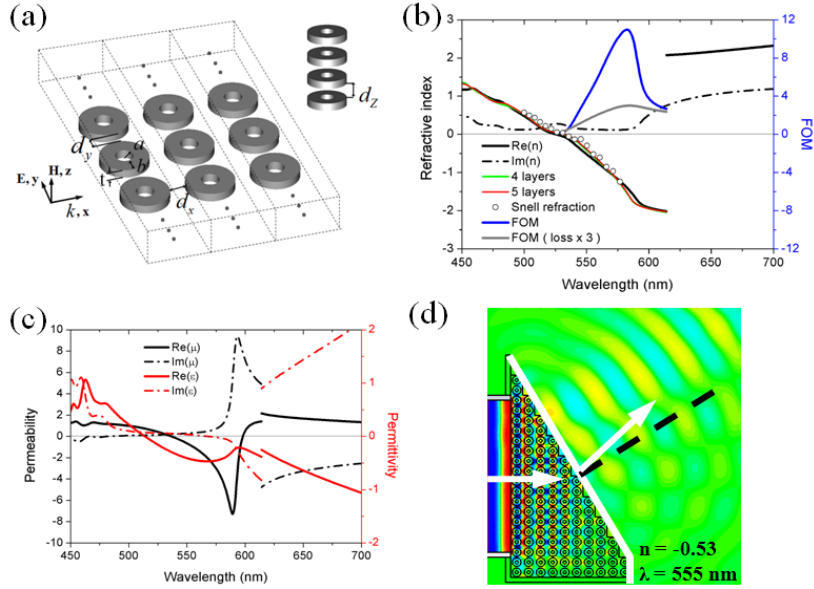

FIG. 1 (color online). (a) The geometries of the proposed NIM of ring structure. The inner radius of the silver rings is $a$ (= 20 nm), the outer radius is $a+b$ (= 55 nm), the thickness is $t$ (= 20 nm), the distances between rings are $d_x$ (= 40 nm), $d_y$ (= 20 nm) and $d_z$ (= 20 nm), in x-, y- and z- directions, respectively. The whole structure is submerged in SiO2 ($\varepsilon = 2.25$). (b) The real and imaginary parts of the effective refractive index and FOM (blue solid line) for 1-layer ring structure. The real part of the refractive index for the 4- (green line) and 5- (red line) layer ring structures converge from that of the 1-layer ring structure. The grey solid line is the FOM when loss increases by a factor of three as compared to that of bulk silver. (c) The real and imaginary parts of the effective permeability and permittivity for 1-layer ring structure. Double negative NIM is realized in the wavelength region from 538 nm to 598 nm. (d) The negative refraction behavior at 555 nm wavelength, indicating an effective refractive index of -0.53.

## 2. Physical mechanism

To study the underlying physical mechanism of the double negative behavior, we plot the distribution of the electric and magnetic fields. In Fig. 2(a), the z-component of the H field at 583 nm (FOM is the highest at this wavelength) indicates a strong magnetic response, which results in a large negative permeability of about -4.7 at this wavelength. Figure 2(b) shows the corresponding electric field. Similarly, the strong electric polarization causes the negative permittivity. The electric field distribution, which is caused by the combination of the magnetic and electric responses, also tells how the magnetic response is formed. We show schematically the magnetic and electric responses in Fig. 2(c) and (d), respectively. For the electric response, the rings behave like electric dipoles, coupled through the gap capacitance $C_E$ between the rings in y direction. For the magnetic response, the rings behave like electric quadrupoles, which alone give little magnetic response. However, when they are closely positioned along y direction, a gap region between two neighboring rings behaves just like a magnetic dipole (with an effective loop current denoted by the red circle with arrows in Fig. 2(c)), giving a very strong magnetic response, which is denoted by $\mathbf{H_r}$. The depolarized field

$H_d$ produced by $H_r$ counteracts the applied external field $H_0$, and consequently may make the direction of the macroscopic field $\overline{H}$ ($\overline{H} = H_0 + H_d$) opposite to $H_0$, resulting in a negative permeability. Furthermore, this magnetic dipole is a bit different from that of SRR or cut-wire pair (a component of the well-known fishnet structure). The charge accumulated at each of the four spots, denoted by "+" or "−" in Fig. 2(c), is due to two oppositely flowing currents. This decreases the current strength required to produce a large magnetic response, and consequently reduces the loss associated to the current in silver. This will also be discussed below with an equivalent LC circuit model.

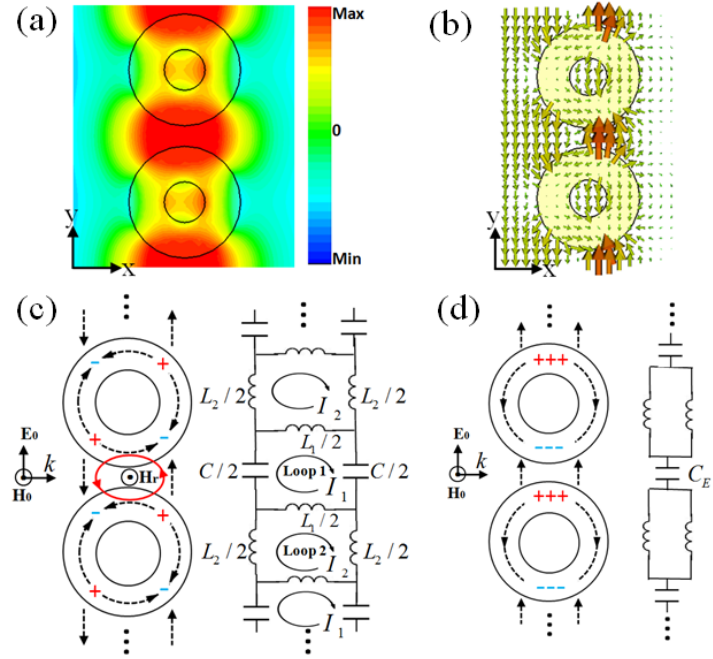

FIG. 2 (color online). The field distribution of (a) $H_z$ and (b) vectorial E field at 583 nm wavelength. (c) For the magnetic response: the left panel shows the schematic E field (dash arrows) and charge distribution (denoted by "+" or "-"), while the right panel shows the equivalent LC circuit model. (d) For the electric response: the left panel shows the schematic E field (dashed arrows) and charge distribution, while the right panel shows the equivalent LC circuit model. For the equivalent LC circuits in (c) and (d), the resistance is not plotted.

Equivalent LC circuits are given in Fig. 2(c) and (d) to provide additional physical understanding of our structure. For loops 1 and 2 (as shown in Fig. 2(c)), we write the equations of Kirchhoff's voltage law (KVL) as

$$I_1(\frac{1}{-i\omega C}) + (I_1 - I_2)(-i\omega L_{1e} + R_1) + I_1(-i\omega L_{m1}) = U_1 = i\omega\mu_0 H_0 S_1, \quad (1)$$

$$I_2(-i\omega L_{2e} + R_2) + (I_2 - I_1)(-i\omega L_{1e} + R_1) + I_2(-i\omega L_{m2}) = U_2 = i\omega\mu_0 H_0 S_2, \quad (1)$$

where $L_{1e}$ ($L_{2e}$) is the total kinetic inductance of the two branches that $L_1$ ($L_2$) resides; $R_1$ ($R_2$) is the total resistance of the two branches where $L_1$ ($L_2$) resides; $L_{m1}$ ($L_{m2}$) is the magnetic inductance of loop 1 (loop 2); $S_1$ ($S_2$) is the area of loop 1 (loop 2); $H_0$ is the applied external field. From the above two equations, we obtain

$$I_1 = \frac{i\omega\mu_0 H_0 \left[ S_1 + \frac{Z_1}{Z_1 + Z_2 + (-i\omega L_{m2})} S_2 \right]}{\frac{1}{-i\omega C} + Z_1 + (-i\omega L_{m1}) - \frac{Z_1^2}{Z_1 + Z_2 + (-i\omega L_{m2})}}, \quad (2)$$

$$I_2 = \frac{i\omega\mu_0 H_0 S_2 + Z_1 I_1}{Z_1 + Z_2 + (-i\omega L_{m2})}, \quad (3)$$

where $Z_1 = -i\omega L_{1e} + R_1$, $Z_2 = -i\omega L_{e2} + R_2$. Then the effective permeability can be derived [19] as

$$\mu_{eff} = \frac{H_0}{H_0 - M} = \frac{H_0}{H_0 - \frac{S_1 I_1 + S_2 I_2}{V}}, \quad (4)$$

where $M$ is the magnetization, $V$ is the volume of the unit cell. Here we estimate according to our structure the values of circuit elements used in Eqs. (1) - (5). We have $Z_1 = l'/(\sigma S')$, where $\sigma = -i\omega(\varepsilon - \varepsilon_0)$ is the AC conductivity, $\varepsilon = -13 + 0.4i$ for the negative permeability in this frequency band, $l' = \pi(a + b/2)$ is the length, $S' = bt$ is the cross section of the wire forming the ring. Thus, the estimation gives $Z_1 = -i\omega L_{1e} + R_1 = -i\omega \cdot 7.14 \times 10^{-13} \text{H} + 70.5\Omega$. Similarly, our estimation gives $Z_2 = -i\omega L_{2e} + R_2 = -i\omega \cdot 7.14 \times 10^{-13} \text{H} + 70.5\Omega$. $S_1$ and $S_2$ are estimated to be $70 \times 120$ (nm$^2$) and $60 \times 110$ (nm$^2$), respectively. Using the method described in [5], $L_{m1}$ and $L_{m2}$ are estimated to be $0.8 \times 10^{-18}$ H, which is much smaller than the kinetic inductance. Finally, capacitance C is estimated to be $1.56 \times 10^{-7}$ pF according to the simulated magnetic resonance frequency. For simplicity and clarity, we have ignored the mutual inductance, since even the self inductance is already one order smaller than the kinetic inductance. This will not influence the validity of our analysis at all. For comparison, we also study the case where only loop 1 exists (in absence of loop 2), which corresponds to the case of a conventional SRR structure or fishnet structure. For this special case, the loop current is

$$I = \frac{i\omega\mu_0 H_0 S_1}{\frac{1}{-i\omega C} + Z_1 + (-i\omega L_{m1})}, \quad (5)$$

and the effective permeability is

$$\mu_{eff} = \frac{H_0}{H_0 - M} = \frac{H_0}{H_0 - \frac{S_1 I}{V}}. \quad (6)$$

We have plotted in Fig. 3 the effective permeability for these two cases. Apparently our structure (black curves in Fig. 3) gives much larger negative permeability and Q factor than the SRR and fishnet structures (red curves in Fig. 3). Comparing $I_1$ (Eq. (3)) with $I$ (Eq. (6)), we noticed that two main factors have contributed to the difference. One lies in the numerator, where the induced electromotive force from loop 2, $U_2 = i\omega\mu_0 H_0 S_2$, also contributes to the loop current of loop 1. The other factor lies in the denominator, where the resistance has been reduced due to the counteracting loop currents $I_1$ and $I_2$ in the branches where $L_1$ resides. Apart from the two main factors related to $I_1$, we noticed from Eq. (5) that $I_2$ also helps make the negative permeability more remarkable. Although Loop 2 alone does not exhibit any resonant behavior without capacitance, it does exhibit strong resonance when coupled with loop 1 as indicated by Eq. (4).

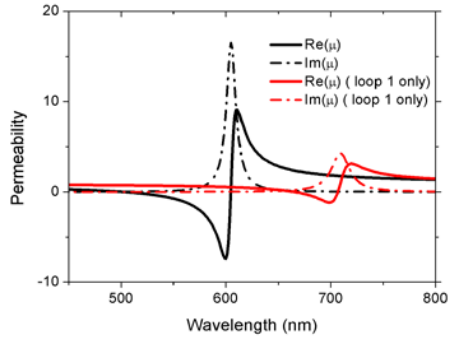

FIG. 3. The real part (solid lines) and imaginary part (dash-dotted lines) of the effective permeability plotted according to Eq. (5) (black lines) and Eq. (7) (red lines).

We carry out additional numerical study to illustrate that the negative refractive index of our structure is a local effect of the magnetic dipole (Loop 1 in Fig. 2) rather than an effect of photonic band structure where the periodicity plays a dominant role. First, we disturb the periodicity in the x direction as shown in the upper inset of Fig. 4(a). When we disturb the periodicity in the x direction, we have kept the distance larger than 30nm so that the coupling between adjacent magnetic dipoles in the x direction won't influence much our result. Apart from the FP resonances, it is clear from Fig. 4(a) that the effective refractive index is almost unchanged when compared with the undisturbed case. Then we disturb the periodicity in the y direction by randomly changing the length of the arms of the rings as shown in the inset of Fig. 4(b) while keeping the magnetic dipole (Loop 1 in Fig. 2) unchanged. Magnetic dipoles are coupled with the neighboring ones in the y direction through the arms of the rings (Loop 2 in Fig. 2), and thus we can significantly change the coupling strength by changing the length of the arms, which in turn change the resonant wavelength. Such resonant wavelength change due to the change of the coupling strength also occurs in SRRs [20]. In Fig. 4(b), the red line is the effective refractive index of the random case (disturbed case), while the green line is the effective refractive index of the periodic case with a fixed arm length of 19.5nm, which is just the average arm length of the random case. Comparing the two cases, we observed only small wavelength shift while the refractive index is almost the same. Thus we have shown that the periodicity in both x and y directions is unnecessary for a negative refractive index of our ring structure, i.e., the negative refractive index of our ring structure is a local effect. This is also true for the disk structure and nanowire structure discussed below.

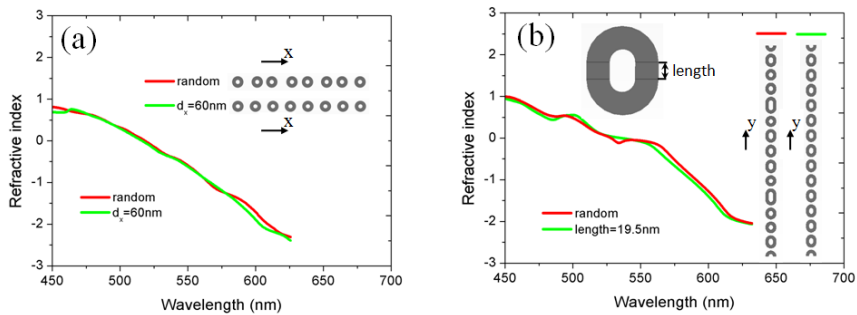

FIG. 4. (a) The effective refractive index after (red line) and before (green line) disturbance of the periodicity in the x direction. (b) The effective refractive index after (red line) and before (green line) disturbance of the periodicity in the y direction.

## 3. Size scaling and structure evolution

Our ring structure can be scaled by varying its geometric parameters to give double negative behavior across the whole visible spectrum and even at UV wavelength. Table 1 shows 15 different sets of geometric parameters. The wavelength region of double negative NIM and the largest FOM for each structure are shown in Fig. 5. This is the first time, to our best knowledge, that double negative NIM at UV wavelength is reported.

For all these structures of different geometric parameters, critical parameter $b$ is always kept not too small, in order to obtain relatively high FOM value. This is partly because $b$ determines the cross section $S' = bt$ of the wire forming the ring, hence the resistance. On the other hand, at very high frequency, the absolute value of the real part of the negative permittivity of silver is much less than that at lower frequency, hence to obtain a strong response to light, we need a larger metal-to-dielectric ratio, which is guaranteed by a large $b$. Alternatively, $a$ is reduced first to push the double negative behavior to shorter wavelength. However, for the convenience of fabrication, $a$ can not be too small either, except zero. Thus when $a$ is required to be less than 15 nm, we simply set $a$ to zero, resulting in a disk-shaped structure (as shown in the second inset of Fig. 5). To reduce the wavelength further, we reduced the kinetic inductance by increasing $t$. Thickness $t$ can not be too large, otherwise the aspect ratio will be too high for fabrication. However, if $t$ is infinitely large (compared with the wavelength), our disk-shaped structure becomes a cylindrical silver nanowire (as shown in the third inset of Fig. 5), which can be electrochemically deposited [21]. By using nanowire structures (samples 10 ~ 15), double negative NIMs at blue-violet and even UV spectrum can be realized with high FOM. We emphasize here that our nanowire structure is double negative, exhibiting negative phase velocity, while the work of [21] is for negative group refractive index (without double negative parameters), which originates from a totally different mechanism.

Table 1. Geometric parameters of the 15 samples.

| Sample # | a (nm) | b (nm) | t (nm) | dz (nm) | dy (nm) |
|---|---|---|---|---|---|
| 1 | 30 | 40 | 20 | 20 | 20 |
| 2 | 25 | 40 | 20 | 20 | 20 |
| 3 | 20 | 40 | 20 | 20 | 20 |
| 4 | 20 | 35 | 20 | 20 | 20 |
| 5 | 15 | 35 | 20 | 20 | 20 |
| 6 | 0 | 50 | 20 | 20 | 20 |
| 7 | 0 | 45 | 20 | 20 | 20 |
| 8 | 0 | 45 | 40 | 20 | 20 |
| 9 | 0 | 45 | 60 | 20 | 20 |
| 10 | 0 | 45 | \ | 0 | 20 |
| 11 | 0 | 40 | \ | 0 | 20 |
| 12 | 0 | 35 | \ | 0 | 20 |
| 13 | 0 | 30 | \ | 0 | 20 |
| 14 | 0 | 30 | \ | 0 | 30 |
| 15 | 0 | 30 | \ | 0 | 40 |

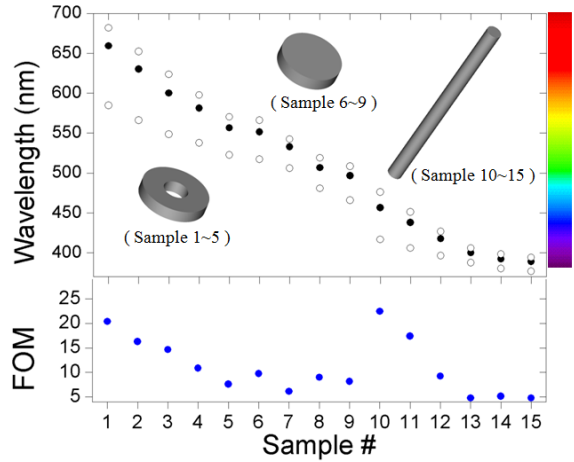

FIG. 5. The double-negative wavelength region (upper plot) and FOM (lower plot) for each sample. For each sample, hollow circles represent the upper bound and lower bound of the double negative region, while black solid circle represents the wavelength where the FOM is the largest. Insets indicate that it is ring structure for samples 1 to 5, disk structure for samples 6 to 9, and nanowire structure for samples 10 to 15.

## 4. Summary

In conclusion, we have proposed three novel types of structures, adapted from the original ring structure, for double negative NIMs across the whole visible spectrum towards UV spectrum with quite high FOM. Among these three types, the nanowire structure is very interesting, due to its capability of exhibiting double negative behavior at even UV spectrum. Furthermore, the extra loss caused by surface roughness can be reduced owing to its small surface-to-volume ratio of the nanowire structure, as compared with layered structures (such as SRRs, fishnet structures, and even our ring and disk structures). Of course, the nanowire structure is limited to 2D applications, while the ring and disk structures are promising for 3D applications after further improvements. We have also analyzed with an equivalent circuit model and explained physically why our novel structure can give double negative behavior with low loss. Our analysis also provides a simple guide for the design of other NIMs of even higher quality at high frequencies.

**Acknowlegments**

We thank Dr. Yongmin Liu for some helpful discussions. This work was partly supported by the National Basic Research Program, the National Natural Science Foundations of China, and the Swedish Research Council (VR) and AOARD.